\documentclass[a4paper,11pt]{article}

\usepackage{amssymb}
\usepackage{graphicx}
\usepackage{dcolumn}
\usepackage{bm}
\usepackage{braket}
\usepackage{amsmath}
\usepackage{amsfonts}
\usepackage{here}
\usepackage{pos}
\usepackage{hyperref}

\let\Re\relax\DeclareMathOperator{\Re}{Re}

\title{Local Polyakov-loop fluctuation and center domains in quark-gluon plasma with many colors}
\ShortTitle{Local Polyakov-loop fluctuation and center domains}

\author*[a]{Yuto Nakajima}
\author[b]{Hideo Suganuma}

\affiliation[a]{Faculty of Science, Kyoto University, \\
Kitashirakawa-oiwake, Sakyo, Kyoto 606-8502, Japan}

\affiliation[b]{Department of Physics, Graduate School of Science, Kyoto University, \\
Kitashirakawa-oiwake, Sakyo, Kyoto 606-8502, Japan}

\emailAdd{nakajima.yuto.62s@st.kyoto-u.ac.jp}
\emailAdd{suganuma@scphys.kyoto-u.ac.jp}

\abstract{
The deconfinement transition in non-Abelian gauge theory is understood as spontaneous breaking of $\mathbb{Z}_N$ symmetry at high temperatures. Accordingly, quark-gluon plasma generally includes some partial cells called center domains, each with a homogeneous Polyakov-loop expectation value.
In this work, constructing an effective action describing the deconfinement vacuum of Yang-Mills theory with $N$ colors, we discuss the properties of center domains.
First, we evaluate the spatial correlation of local Polyakov-loop fluctuation and demonstrate that some fluctuation becomes a Nambu-Goldstone-like mode in the large-$N$ limit. We also discuss surface tension between two $\mathbb{Z}_N$ center domains. Second, we estimate the global vacuum-to-vacuum transition in a single center domain. We find that some threshold volume exists, where a domain larger than this volume is stable, and vice versa. Identifying the threshold as the lower bound of a stable center domain volume, we quantitatively argue the typical volume scale of center domains.
}

\FullConference{%
The 40th International Symposium on Lattice Field Theory (Lattice2023), \\
31 July-4 August, 2023\\
Chicago, USA
}

\begin{document}
\maketitle

\section{Introduction}

Color confinement, a distinctive feature of quantum chromodynamics (QCD), ceases to exist under high temperatures. Both theoretical and experimental physicists have extensively investigated the deconfined system with keen interest. 
From a symmetry standpoint, the deconfinement is explained as spontaneous symmetry breaking (SSB) of the global $\mathbb{Z}_N$ symmetry, which corresponds to the center of the gauge group SU($N$)~\cite{1975Polyakov}. 
This results in a nontrivial $\mathbb{Z}_N$ structure in the deconfinement Yang-Mills vacuum due to its spontaneous breaking.

This study aims to examine the deconfinement vacuum structure utilizing the Polyakov-loop effective model of the SU($N$) Yang-Mills theory. 
We construct an effective model that includes the traced Polyakov-loop field $\phi(\boldsymbol{x})$ and investigate its fluctuation properties beyond a spatially uniform vacuum.
Our initial objective revolves around a quantitative assessment of the correlation length of the Polyakov loop phase $\mathrm{Arg}\left( \phi(\boldsymbol{x})\right)$. 
At elevated temperatures, the theory exhibits $N$ different degenerate vacua, reflective of the spontaneously broken $\mathbb{Z}_N$ symmetry.
In the large-$N$ limit, we expect the potential wall between these vacua to vanish, transforming the global $\mathbb{Z}_N$ symmetry into an approximately continuous U(1) symmetry. 
Consequently, the massive mode is reduced into a massless mode, akin to an extension of the Nambu-Goldstone theorem.

\begin{figure}[thbp]
\centering
\includegraphics[width=150mm]{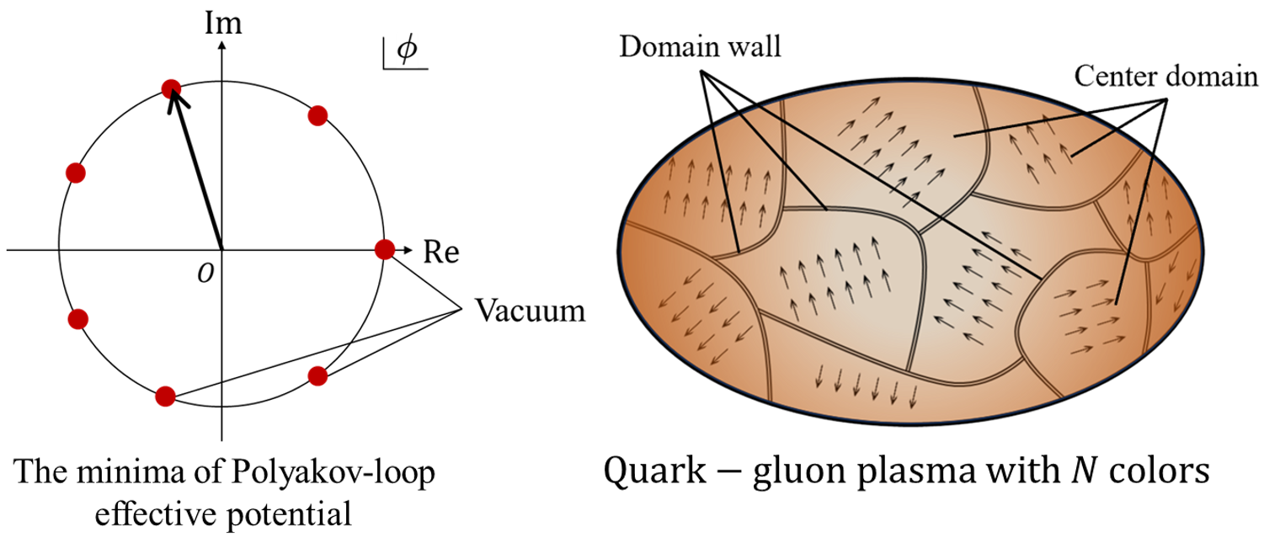}
\caption{The Polyakov-loop effective potential and the center domain of the quark-gluon plasma. 
The arrows symbolize the phases of the vacuum expectation value of the Polyakov loop. 
Each domain in quark-gluon plasma, separated by potential walls, is characterized by its Polyakov-loop configuration.}
\label{domainwall}
\end{figure}

This paper also explores the typical volume of center domains within a quenched quark-gluon plasma. 
The system is envisioned to be divided into numerous small-volume center domains, as depicted in Fig.~\ref{domainwall}. 
Each domain is characterized by its Polyakov-loop expectation value $\braket{L}$, representing one of the $N$ degenerate vacua. 
We estimate the timescale for the domains to persist in one of the potential minima, elucidating how their volumes are evaluated from the perspective of stability.

\section{Formulation and Symmetry}
\label{sec4}

Our formulation begins from the lattice action for SU($N$) Yang-Mills theory
\begin{align}
S_{\mathrm{YM}}[U] 
= -\frac{2}{g^2} \sum_{s}
\sum_{\mu < \nu}
\Re \mathrm{Tr}\ \square_{\mu \nu}(s).
\label{plaquette}
\end{align}
The thermal average of the Polyakov loop
\begin{align}
L_i = \prod_{i_{\tau}=1}^{N_{\tau}} U_{\tau} (i_{\tau}, i) \in {\rm SU}(N)
\label{polyakovloop}
\end{align}
serves as an order parameter characterizing the confinement.

The action (\ref{plaquette}) is invariant under the spatially global $\mathbb{Z}_N$ transformation: $U_{\tau} (\tau_0, i) \rightarrow z U_{\tau} (\tau_0, i)$ $(z = e^{i \frac{2\pi n}{N}}, n = 0, 1, ... , N-1)$, under which the Polyakov loop (\ref{polyakovloop}) changes as  $L_i \rightarrow z L_i$.
An effective action that includes the Polyakov loop as its dynamical variable is expected to take over the original $\mathbb{Z}_N$ invariance, and the symmetry undergoes spontaneous breaking at high temperatures. 
Employing a strong coupling approximation~\cite{1982Polonyi}, we integrate out all space-like link variables, resulting in
\begin{align}
    Z_{\mathrm{YM}} 
  =
    \int \left( \prod_{i_{\tau}=1}^{N_{\tau}} \prod_{i}\mathrm{d} U_{\tau}(i_{\tau}, i) \right)  \exp \left[\lambda_{\mathrm{YM}}^{-N_{\tau}}\sum_{\braket{i, j}} \mathrm{Tr}( L_i ^{\dagger}) \ \mathrm{Tr}( L_j )\right]
    \label{ZYM00}
    \end{align}
up to the leading order, where ’t~Hooft coupling is denoted as $\lambda_{\mathrm{YM}}=g^2N$. 
The summation $\Sigma_{\braket{i,j}}$ covers all nearest-neighbor variables. 
The effect of the SU($N$) group integration can be taken into account by expressing the Haar measure as part of an effective potential. 
Then, one finds 
\begin{align}
Z_{\mathrm{YM}} 
  =
  \int \mathcal{D}\phi \exp 
\left[ - \phi^{\dagger} \hat{\mathcal{J}}^{-1} \phi
 + \sum_i \ln \mathcal{H}^{(N)}\right]
 \quad
\left( \phi_i \equiv \frac{1}{N} \mathrm{Tr}(L_i) \right).
\label{ZYM}
\end{align}
Here $\phi = (\phi_1, \phi_2, ...)$, and the string tension at zero temperature is given as $\sigma \equiv a^{-2} \ln \lambda_{\mathrm{YM}}$. 
We can write the Haar measure $\mathcal{H}^{(N)}$ as a function of the traced Polyakov loop $\phi_i$ by taking the Polyakov gauge.
See \cite{Nakajima:2023} for further information including the explicit form of the kernel $\hat{\mathcal{J}}^{-1}$.

Coarse graining of the short-wavelength modes results in the effective action
\begin{align}
S_{\mathrm{YM}}[\phi(\boldsymbol{x})]
=
C
\int \mathrm{d}^3 \boldsymbol{x}
\Bigg(
|\nabla \phi(\boldsymbol{x})|^2
-
\frac{6}{a^2}
|\phi(\boldsymbol{x})|^2
-
\frac{e^{\sigma a/T}}{N^2 a^2}
\ln \mathcal{H}^{(N)} (\phi(\boldsymbol{x}))
\Bigg).
\label{SYM2}
\end{align}
The corresponding partition function is given by $Z_{\mathrm{YM}} = \int \mathcal{D} \phi \ e^{-S_{\mathrm{YM}}[\phi(\boldsymbol{x})]}$, where $C$ is a constant. 
Although this model is constructed on a lattice, it can be treated as a continuous field model, focusing specifically on long-range correlations in the infrared region where $|\boldsymbol{k}| \ll a^{-1}$.

Next, we proceed to a crucial component in (\ref{SYM2}): the effective potential $\ln \mathcal{H}^{(N)}(\phi)$, which is invariant under the global $\mathbb{Z}_N$ transformation $\phi \mapsto e^{i \frac{2 \pi n}{N}} \phi$. 
Since the effective potential for SU(2) and one for SU(3) are exactly known~\cite{FUKUSHIMA2017154}, our focus now shifts to examining the scenario for $N \geq 4$.
In the absence of a known explicit form for these cases, we opt for the simplest form that preserves $\mathbb{Z}_N$ symmetry in our model, aligning with Sannino's proposal~\cite{2005Sannino}
\begin{align}
\mathcal{H}^{(N)}(\phi) = 1- b_2 |\phi|^2 - b_4 |\phi|^4 + b_N \mathrm{Re}\ \phi^N.
\label{HN}
\end{align}
As a caution, this form may not be suitable for analyses on the deconfinement phase transition, for (\ref{HN}) shows second-order transitions for even $N$'s, whereas they are expected to be first-order ones. 
Nevertheless, we see no issues as long as the discussion is limited to the deconfinement phase.

As shown below, the effective action (\ref{SYM2}) suggests the emergence of a massless mode in the large-$N$ limit \cite{Nakajima:2023}. 
We delve into the fluctuations around one of the potential minima and examine the angle $\theta(\boldsymbol{x})$ correlation function.
Assuming a frozen amplitude i.e. $\phi(\boldsymbol{x}) = l_0 e^{i \theta(\boldsymbol{x})/l_0}$ with $l_0 \equiv \langle |\phi (\boldsymbol{x})| \rangle \in (0,1)$, we expand (\ref{SYM2}) in terms of $\theta(\boldsymbol{x})$ to acquire
\begin{align}
    S_{\mathrm{YM}}
    &\sim
    2C\int \mathrm{d}^3 \boldsymbol{x}
\left[
\frac{1}{2}
(\nabla \theta)^2 + V_{\mathrm{YM}}(\theta) 
\right],
\label{SYM3}
\end{align}
where $b \equiv 1-b_2 l_0^2 - b_4 l_0^4 + b_N l_0^N$ and
 \begin{align}
V_{\mathrm{YM}}(\theta) \equiv
\frac{m^2_{\mathrm{YM}}l_0^2}{N^2}
\left( 1 - \cos \left( \frac{N}{l_0}\theta \right)\right), \quad
m_{\mathrm{YM}} \equiv \sqrt{\frac{e^{ \sigma a/T}}{2 a^2}
\frac{b_N l_0^{N-2}}{ b}}.
\label{mYM}
\end{align}
For a small $\theta$ fluctuation around $\theta = 0$, the correlation function of $\theta(\boldsymbol{x})$ is given by
\begin{align}
\braket{\theta(\boldsymbol{x}) \theta(0)}
=
\frac{\int \mathcal{D}\theta \ \theta(\boldsymbol{x}) \ \theta(0) \ e^{-S_{\mathrm{YM}}}}
{\int \mathcal{D}\theta e^{-S_{\mathrm{YM}}}} \propto
\frac{1}{|\boldsymbol{x}|} e^{-m_{\mathrm{YM}} |\boldsymbol{x}|}.
\end{align}
This mode exhibits a Yukawa-type spatial correlation with a range of $m_{\mathrm{YM}}^{-1}$, where $m_{\mathrm{YM}}$ is called the screening mass. 
Considering (\ref{mYM}) and the upper bound constraint $l_0<1$, we conclude
\begin{align}
    \lim_{N \rightarrow \infty} m_{\mathrm{YM}} = 0
\end{align}
in the large-$N$ limit. 
A Coulomb-type spatial correlation with an infinite range emerges, akin to a Nambu-Goldstone mode.

Moreover, we can calculate the surface tension $\alpha$ between the adjacent domains using the parameters shown in (\ref{SYM3}). 
The surface tension for the model with cosine-type potential is analytically known~\cite{Monden:1998}, and we obtain 
\begin{align}
    \alpha = \frac{N}{2a^3}\frac{b_N}{b}l_0^N.
\end{align}
This tension diminishes in the large-$N$ limit, indicating the vanishing of the potential barrier between adjacent vacua in this limit.

Both of these outcomes give us circumstantial evidence that a Nambu-Goldstone-like mode emerges in this model. 
In other words, these results suggest that under the large-$N$ limit the $\mathbb{Z}_N$-symmetric potential is transformed into a U(1)-symmetric one, and the structure of symmetry of the action is qualitatively changed.

\section{Center Domain Volume} 
\label{sec5}

\begin{figure*}[htbp]
  \begin{minipage}[b]{0.48\linewidth}
      \centering
  \includegraphics[width=70mm]{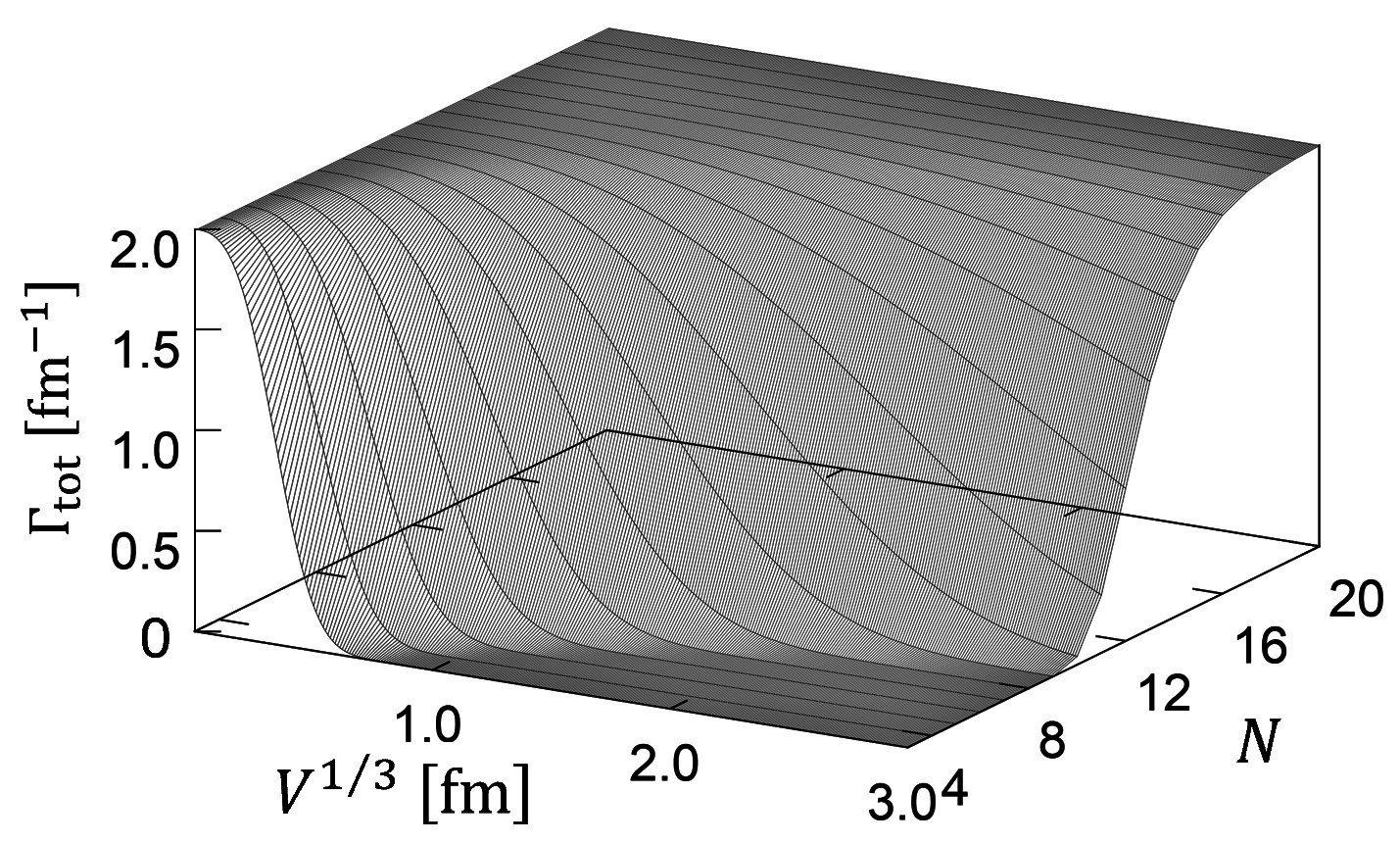}
  \end{minipage} %
  \begin{minipage}[b]{0.48\linewidth}
      \centering
  \includegraphics[width=70mm]{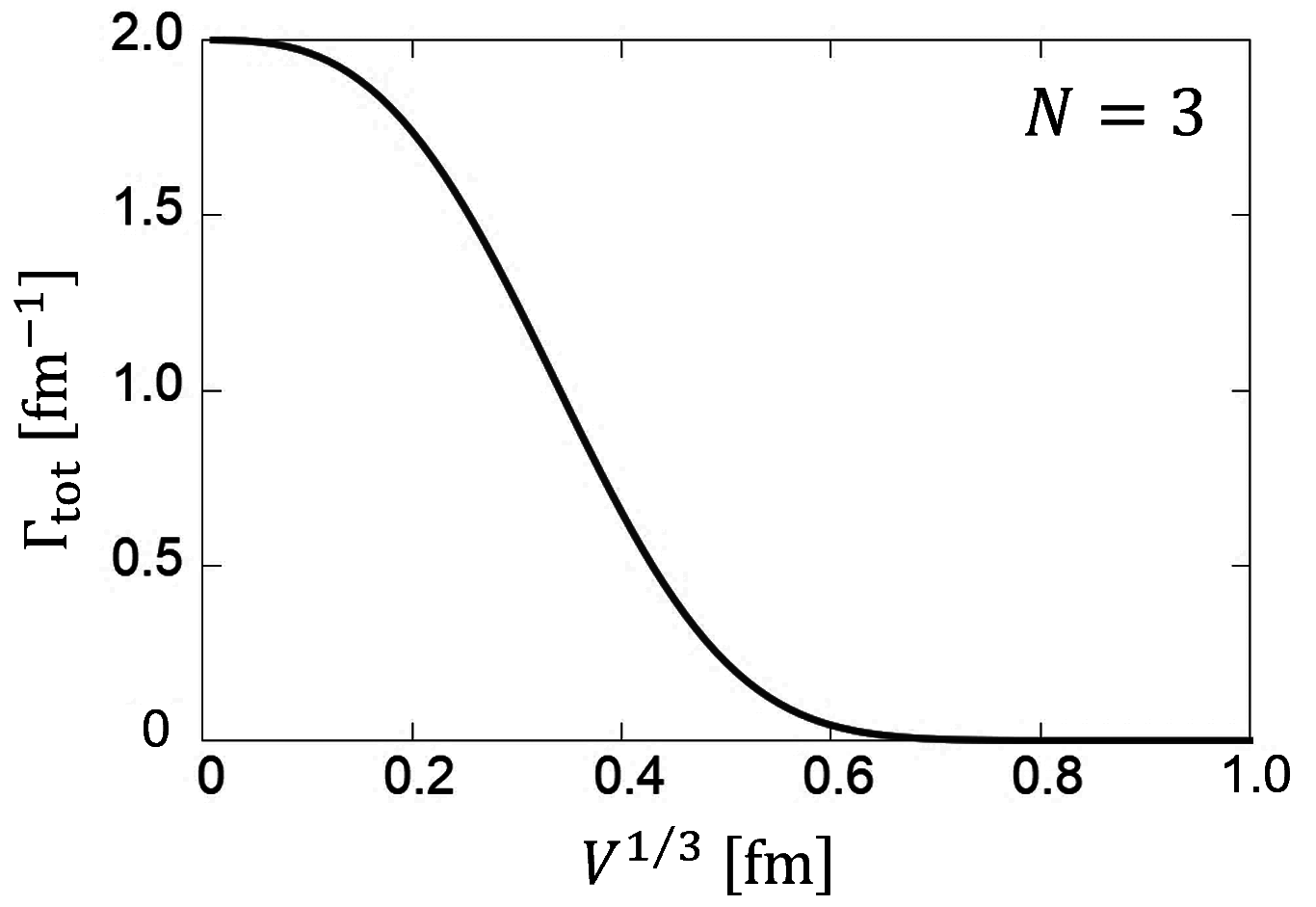}
  \end{minipage}
  \centering
  \caption{
The total transition rate of the domain as a function of $V^{1/3}$ and $N$ in the left panel, while the right panel focuses on the $N=3$ case. 
These rates are computed using the following parameter set: lattice spacing $a = 0.4 \ \mathrm{fm}$, temperature $T = 400 \ \mathrm{MeV}$, string tension at zero temperature $\sigma = 1.0 \ \mathrm{GeV/fm}$, the vacuum expectation value $l_0 = 0.5$, and $b_N/b = 7.52$ (the value when $N=3$).
Taken from \cite{Nakajima:2023}.}
  \label{grapf_V,N}
\end{figure*}

 \begin{figure*}[htbp]
  \begin{minipage}[b]{0.48\linewidth}%
      \centering
  \includegraphics[width=70mm]{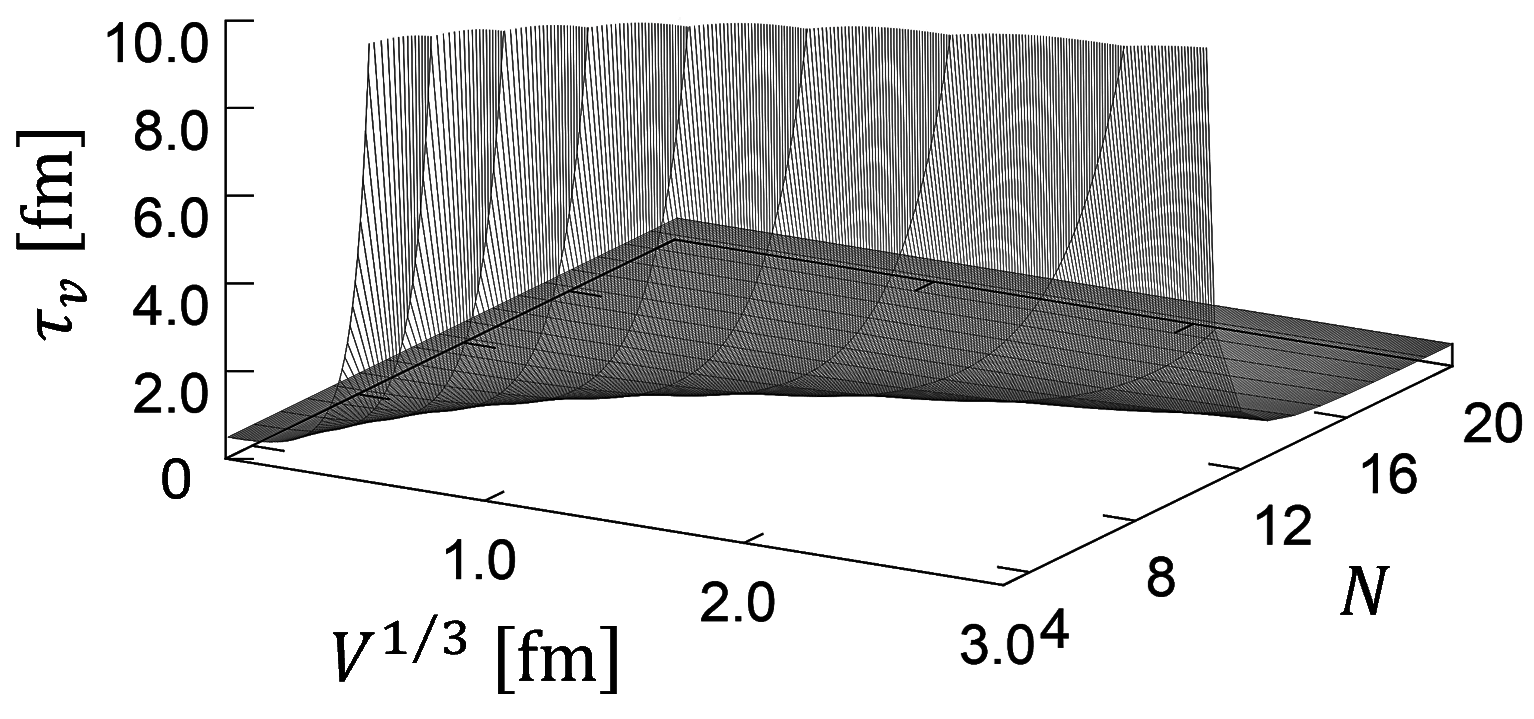}
  \label{lifetime}
  \end{minipage}%
  \begin{minipage}[b]{0.48\linewidth}%
      \centering
  \includegraphics[width=70mm]{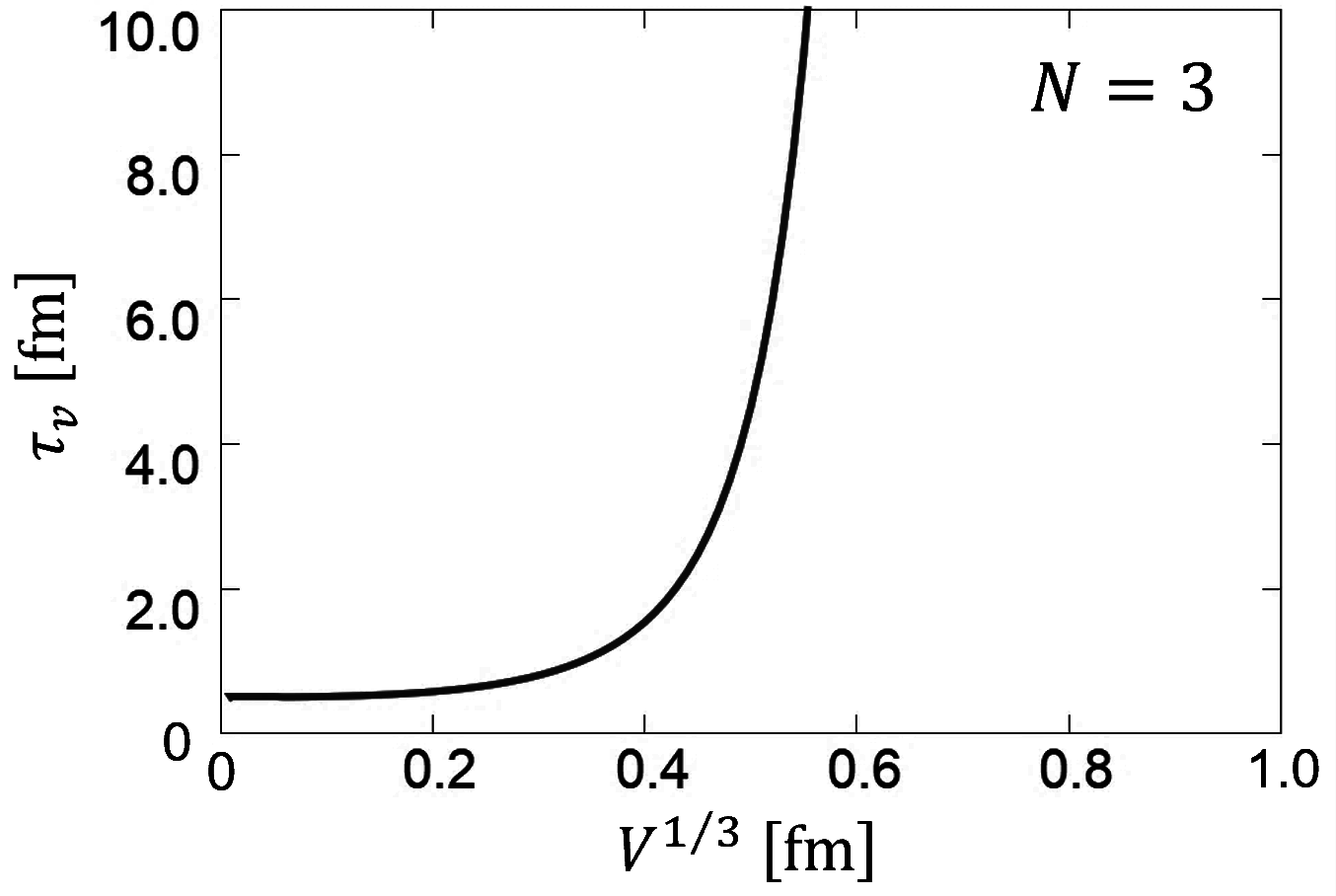}
  \end{minipage}
  \centering
  \caption{
  The lifetime of the domain as a function of $V^{1/3}$ and $N$ in the left panel, with a focus on the case where $N=3$ in the right panel. 
  Here the same set of parameters as in Fig.~\ref{grapf_V,N} are used. Taken from \cite{Nakajima:2023}.}
  \label{grapf2_V,N}
\end{figure*}

Next, we proceed to estimate the lifetime of a domain and the volume of the center domain based on (\ref{SYM3}). 
To simplify the analysis, we assume that the dominant path occurs along the circumference $|\phi|=l_0$, reducing the transition on the $\phi$-plane to a one-dimensional system.

In the subsequent discussion, we assume that the field $\theta(\boldsymbol{x})$ exhibits an imaginary-time dependence, and specifically, that $\theta(\tau, \boldsymbol{x})$ is homogeneous concerning the spatial coordinate $\boldsymbol{x}$:
\begin{align}
S_{\mathrm{QM}} 
=
\int \mathrm{d}\tau 
\Bigg[\frac{M}{2} \left(\frac{\partial \theta}{\partial \tau}  \right)^2
+ \frac{V_0}{2}
\left( 1 - \cos \left( \frac{N}{l_0} \theta \right)\right)\Bigg],
\end{align}
where
\begin{align}
M(V,N) \equiv \left( \frac{N}{a} \right)^2 \cdot 2V e^{ \sigma a/T},\quad
V_0(V,N) \equiv  \frac{2V}{a^4} \frac{b_N}{b} l_0^N
\label{V_0}
\end{align}
It is noteworthy that in (\ref{V_0}) the $N$ dependence manifests in $b_N/b$ and $l_0^N$, with the dominant contribution arising from the latter exponentiation. 
The action can be interpreted as that of a ($1+0$)-dimensional quantum mechanical system, and we understand the vacuum-to-vacuum transition as the dynamics of a virtual particle following this action.

The transition between two adjacent potential minima can take place through both thermal transitions and tunneling processes. 
The total transition rate per unit of time, denoted as $\Gamma_{\mathrm{tot}}(V, N)$, is expressed as the sum of these two rates
\begin{align}
\Gamma_{\mathrm{tot}}(V, N)  =\Gamma_{\mathrm{th}} (V, N) 
+\braket{\Gamma_{\mathrm{tun}}(V, N; E) }.
\label{Gtot}
\end{align}
The first term is obtained on the assumption that the trapped virtual particles have energy following the canonical distribution, and particles with energy above the potential depth can surpass the wall.
The second term, the tunneling rate, is calculated based on WKB approximation, and $\braket{\bullet}$ represents the thermal expectation value.
Consequently, we obtain the transition rates $\Gamma_{\mathrm{tot}}(V,N)$ analytically and use them to estimate the lifetime of a domain, denoted as $\tau_v(V,N) = 1/\Gamma_{\mathrm{tot}}(V,N)$.
See~\cite{Nakajima:2023} for further information.

\begin{figure}[thbp]
\centering
\includegraphics[width=100mm]{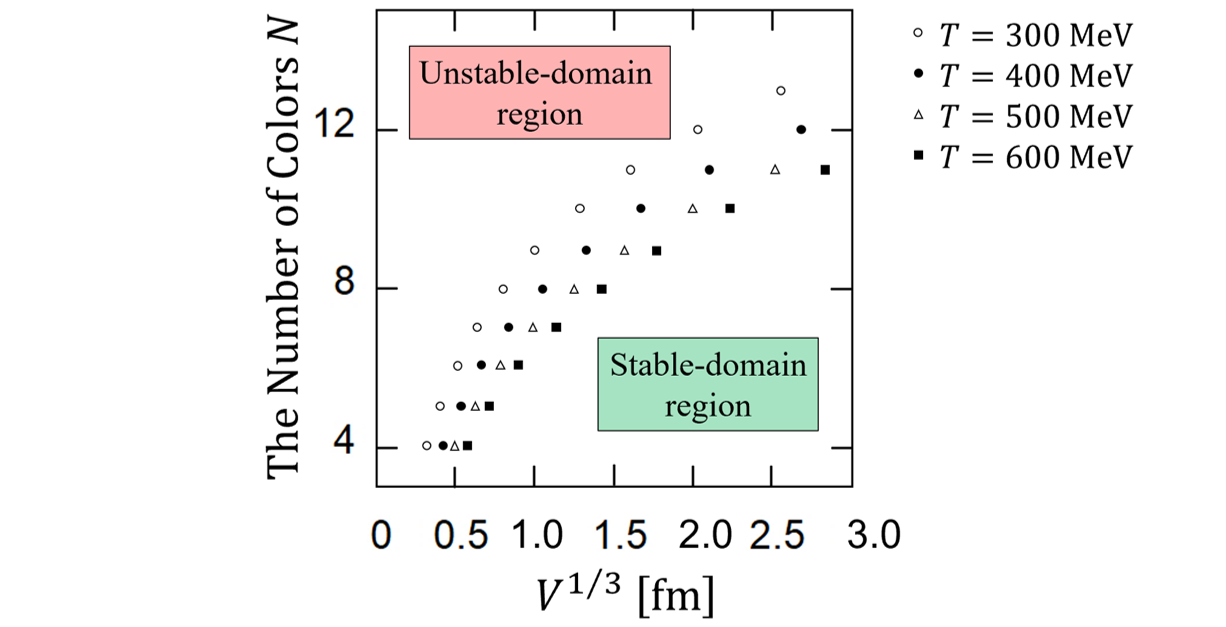}
\caption{
The fuzzy boundaries between the unstable-domain and stable-domain regions are depicted for various temperatures: $T$ = 300MeV, 400MeV, 500MeV, and 600MeV. 
On the transition curves, the domain's lifetime extends to $1.0 \ \mathrm{fm}$, corresponding to the typical timescale of hadrons. 
Domains smaller than the transition thresholds, or unstable domains, are anticipated to shrink and vanish, whereas those surpassing the thresholds are expected to be stabilized.
Taken from~\cite{Nakajima:2023}.}
\label{intercept}
\end{figure}

Figure~\ref{grapf_V,N} illustrates the dependence of the total transition rate $\Gamma_{\mathrm{tot}}$ on $(V, N)$. 
Along a specific curve, the transition rate experiences a sharp decrease. 
The corresponding lifetime $\tau_v$ is depicted in Fig.~\ref{grapf2_V,N}.
This figure reveals that the $(V^{1/3}, N)$-plane divides into two different regions: the stable and unstable-domain regions. 
The boundary between these two phases is a fuzzy crossover, defined by the rapid increase in lifetime. 
The "critical curve," where the lifetime intersects 1.0 $\mathrm{fm}$, is presented in Fig.~\ref{intercept} for $T=$ 300MeV, 400MeV, 500MeV, and 600MeV.

In high-energy heavy-ion collision experiments, the system undergoes division into numerous small center domains, each characterized by a different Polyakov-loop configuration. 
Since domains smaller than the "critical volume" tend to decay, the typical lower bound of center domain volumes aligns with this "critical volume."

\section{Conclusion}
\label{sec6}

In this paper, we have analytically modeled the non-trivial $\mathbb{Z}_N$ structure of the deconfinement phase in SU($N$) Yang-Mills theory and obtained an interesting depiction of the dynamics of center domains based on Polyakov-loop effective action.

In the initial part, we developed the Polyakov-loop effective action for the theory with a finite number of colors. 
We explored the correlation function of the Polyakov loop and derived the fluctuation mass as a function of the color number based on the model.
The analysis also encompassed the metamorphosis of the global symmetry structure in the large-$N$ limit, with a focus on the fluctuation of the Polyakov loop along its angle direction and the surface tension between adjacent domains. The results affirm that this limit transforms the $\mathbb{Z}_N$-symmetric theory into a U(1)-symmetric theory.

In the subsequent part, we delved into the global $\mathbb{Z}_N$ structure in the finite-volume quark-gluon plasma. 
Our study focused on the typical volume scale of one of the center domains in the quark-gluon plasma as a function of the color number and volume. 
The findings revealed that the $(V^{1/3}, N)$-plane can be divided into two regions: stable-domain and unstable-domain regions. 
This indicates that a center domain is stable if its volume exceeds a certain threshold, while a center domain with a volume below it is unstable.

As a technical improvement, beyond the strong coupling expansion in this paper, it is important and desired to perform lattice QCD Monte-Carlo simulations to numerically verify our results.

\section*{Acknowledgment}

H.S. is supported in part by the Grants-in-Aid for
Scientific Research [19K03869] from Japan Society for the Promotion of Science.

\bibliographystyle{JHEP} 
\bibliography{main_1017.bib}

\providecommand{\noopsort}[1]{}\providecommand{\singleletter}[1]{#1}%

\providecommand{\href}[2]{#2}\begingroup\raggedright\begin{thebibliography}{1}

\bibitem{1975Polyakov}
A.~Polyakov, \emph{Compact gauge fields and the infrared catastrophe},
  \href{https://doi.org/https://doi.org/10.1016/0370-2693(75)90162-8}{\emph{Phys.
  Lett. B} {\bfseries 59} (1975) 82}.

\bibitem{1982Polonyi}
J.~Polonyi and K.~Szlachanyi, \emph{Phase transition from strong-coupling
  expansion},
  \href{https://doi.org/https://doi.org/10.1016/0370-2693(82)91280-1}{\emph{Phys.
  Lett. B} {\bfseries 110} (1982) 395}.

\bibitem{Nakajima:2023}
Y.~Nakajima and H.~Suganuma, \emph{{$\mathbb{Z}_N$} structure of deconfinement
  vacuum in {SU($N$}) {Y}ang-{M}ills theory: emergence of {N}ambu-{G}oldstone
  mode in large-{$N$} limit},  2023,
  \href{https://arxiv.org/abs/2212.13874}{{\ttfamily
  arXiv:2212.13874[hep-th]}}.

\bibitem{FUKUSHIMA2017154}
K.~Fukushima and V.~Skokov, \emph{Polyakov loop modeling for hot {QCD}},
  \href{https://doi.org/https://doi.org/10.1016/j.ppnp.2017.05.002}{\emph{Prog.
  Part. Nucl. Phys.} {\bfseries 96} (2017) 154}.

\bibitem{2005Sannino}
F.~Sannino, \emph{Higher representations: Confinement and large{$N$}},
  \href{https://doi.org/10.1103/PhysRevD.72.125006}{\emph{Phys. Rev. D}
  {\bfseries 72} (2005) 125006}.

\bibitem{Monden:1998}
H.~Monden, H.~Ichie, H.~Suganuma and H.~Toki, \emph{Surface tension in the
  {QCD} phase transition in the dual {G}inzburg-{L}andau theory},
  \href{https://doi.org/10.1103/PhysRevC.57.2564}{\emph{Phys. Rev. C}
  {\bfseries 57} (1998) 2564}.

\end{thebibliography}\endgroup

\end{document}